\documentclass[epj]{svjour}
\usepackage{graphics}
\usepackage{amsmath,amssymb}
\usepackage[colorlinks,citecolor=blue,linkcolor=blue]{hyperref}

\begin{document}

\title{Quantum transport of double quantum dots coupled to an oscillator in arbitrary strong coupling regime}
\author{Chen Wang\inst{1,2} \and Jie Ren\inst{2,3,*} \and Baowen Li\inst{2,3,\dag} \and Qing-Hu Chen\inst{4,1,\ddag}}
\institute{Department of Physics,
Zhejiang University, Hangzhou 310027, P. R. China \and
Department of Physics and Centre for Computational
Science and Engineering, National University of Singapore, Singapore
117546, Republic of Singapore \and NUS
Graduate School for Integrative Sciences and Engineering, Singapore
117456, Republic of Singapore \and Center for Statistical and
Theoretical Condensed Matter Physics, Zhejiang Normal University,
Jinhua 321004, P. R. China\\
\email{$^*$phyrj@nus.edu.sg; $^{\dag}$phylibw@nus.edu.sg; $^{\ddag}$qhchen@zju.edu.cn}
}

\date{Received: date / Revised version: date}

\abstract{
In this paper, we investigate the quantum transport
of a double quantum dot coupled
with a nanomechanical resonator at arbitrary strong electron-phonon
coupling regimes. We employ the generalized quantum master equation
to study full counting statistics of currents. We demonstrate
the coherent phonon states method can be applied to decouple the electron-phonon interaction non-perturbatively. With the help of this non-perturbative treatment of electron-phonon couplings,
we find that the phonon-assisted resonant tunneling emerges
when the excess energy from the left quantum dot to the right one can
excite integer number of phonons and multi-phonon excitations can enhance the transport in strong electron-phonon coupling regime.
Moreover, we find that as the electron-phonon coupling increases, it first plays a constructive role to assist the transport,
and then plays the role of scattering and strongly represses the transport.
}


\authorrunning{}
\titlerunning{EPJB}

\maketitle

\section{Introduction}
With the increasing promotion of nanotechnology, people now have the
abilities to fabricate fertile atomic and molecular quantum devices,
such as single superconducting electron transistors~\cite{Fulton},
quantum dots~\cite{Ashoori} and single molecular
transistors~\cite{SMT}. As a consequence, great attentions have
recently been attracted on research fields of molecular
electronics~\cite{Aviram,Joachim,Bumm,Reed} and
nanoelectromechanical systems~\cite{Craighead,Cleland,Blencowe}, because of their
immense potentials to bridge scientific researches and industry
applications. One of such promising devices is Double Quantum Dot
(DQD) system~\cite{Jong1}, which, as artificial atoms, is able to
confine one or several electrons to form effective two level system,
so as to manipulate and control coherent transport of qubits by
superposition of individual electrons' natural states
~\cite{Nakamura,Vion,Chiorescu,Yu}.

It was initially observed by Fujisawa \emph{et al}.
experimentally~\cite{Fujisawa} and then explained by Brandes
\emph{et al}. theoretically~\cite{Brandes1} that single electron
tunneling on such quantum dots system inevitably interplays with
its environment, such as electron-phonon (e-ph) interaction due to vibrations of the system. The transport properties are extremely sensitive to the motion of the nanomechanical resonator such that they can be also utilized, in a reverse way, as detectors with high precisions for the quantization of the resonator
positions~\cite{Blencowe,Knoibel,Lahaye,Naik,Clerk}.
Therefore, it is definitely important
to unravel the underlying interplay effect between electronics and
vibrations on transport properties for future quantum technology
development.

In the other hand, the main transport property, current, has been extensively studied in the past decades. However, current noise is not yet fully exploited as a
powerful tool to extract more information to understand how the
environment affects the transport~\cite{GK1,Sanchez2008}. Especially, at
low temperatures, shot noise~\cite{Blanter} is the main source
contributing to current fluctuations because of the discreteness of
transferring qubits and charges. Full counting statistics (FCS)~\cite{Nazarov2}, initially proposed by Levitov \emph{et al}.~\cite{Levitov1} in mesoscopic physics, is
proved as a splendid diagnostic tool to investigate complete
information of quantum fluctuations both in
calculations~\cite{Sun,bargrets1,Aguado1,bargrets2,Ren} and
measurements~\cite{Wei1,Bylander,Gustavsson}. The main idea of FCS
is, by introducing auxiliary counting parameters, to evaluate the
cumulant generating function of probability distribution of
transmitted charges. Current and shot noise are just the first two
cumulants. The higher order cumulants indeed show
more details to expand our views to study quantum fluctuation
effects~\cite{petter2011,Flindt,Flindt2}. The underlying mathematics of FCS is
named large deviation theory~\cite{LDT1}, which has been widely used
in various fields~\cite{LDT2,LDT3}.

In this paper, we focus on effects of quantized vibrations of
the nanomechanical resonator on transport
properties of the DQD with large voltage bias between two
electrodes. The vibration is specified by a single phonon mode, which interacts with an external thermal phonon environment. We derive a generalized quantum master equation (QME) including auxiliary counting parameters under the Born-Markov approximation
and second-order perturbation of system-bath couplings.
Though two-level system strongly coupled to the oscillator
was studied intensively, people usually applied rotating wave approximation (RWA)
or only investigated the ground state by variational methods in the closed systems~\cite{irish1,hwang1}.
While extended coherent phonon states method~\cite{Chen1,Chen2} is adopted to capture the characteristics of excitations mixed by electrons and phonons with
arbitrary strong e-ph coupling in the absence of RWA. This
non-perturbative approach renders us beyond
the previous perturbation expansion method by assuming a weak e-ph coupling~\cite{Lambert1}. 
In fact, the strong e-ph interaction has already been found in carbon based
nanoscale devices, shown in Refs.~\cite{strongep1,strongep2,strongep3}. In recent quantum electrodynamics experiments, the coupling of the two-level artificial atom and the resonator also reaches the strong regime~\cite{strongep4}. 
Therefore, our work about coherent phonon states approach, which is capable of dealing with arbitrary e-ph
coupling strengths, is important and favorite.

The organization of this paper is as follows: we begin by modeling
the system with generalized QME combined with coherent phonon states
approach in the framework of FCS in section II. Details about how
coherent phonon states can optimally decouple e-ph interaction
non-perturbatively will be given there. In section III, results and
discussions are presented, of which the first three cumulants of
probability distribution of transmitted charges are scrutinized in
the full range of e-ph coupling strengths. Finally, summary is given in
section IV.

\section{Theory and Methods}

In this section, we first describe the whole Hamiltonian and then
give the derivation of the generalized QME accompanied by auxiliary
counting parameter and FCS. Finally, extended coherent phonon states approach
will be articulated to show its ability to decouple the e-ph
interaction.

\subsection{The Whole Hamiltonian}

\begin{figure}
\resizebox{0.45\textwidth}{!}{\includegraphics{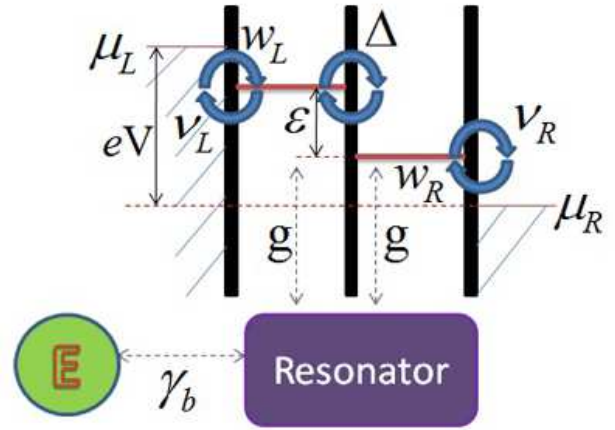}}\vspace{0.5cm}
\caption{ (Color online) Sketch map of the DQD (among three vertical
thick lines, specified by two horizontal lines) coupled with the
nanomechanical resonator (rectangle). Left and right areas with
slash lines denote electrodes. The thermal environment of phonons is
depicted by E (circle).}\label{fig:fig1}
\end{figure}

Hamiltonian of the whole system, as illustrated in
Fig.~\ref{fig:fig1}, is generally described by
\begin{eqnarray}
H=H_S+H_{\mathrm{Lead}}+H_{\mathrm{E}}+V_{\mathrm{DL}}+V_{\mathrm{RE}},
\end{eqnarray}
where $H_S$ denotes the coupled system of DQD and nanomechanical
resonator. $H_{\mathrm{Lead}}$ depicts the two electrodes.
$H_{\mathrm{E}}$ is the environment. $V_{\mathrm{DL}}$ describes the
coupling between DQD and leads, while $V_{\mathrm{RE}}$ denotes the
coupling between resonator and thermal environment. Their detailed
expressions are to be explained in the following.

We assume only one electron is allowed to stay in DQD at most,
considering strong coulomb repulsion in quantum dots. The system of
the DQD coupled with a quantized nanomechanical resonator thus is
expressed as:
\begin{eqnarray}~\label{Hs}
H_S=\frac{\varepsilon}{2}\sigma_z+\Delta\sigma_x+g\sigma_z(a^\dag+a)+\omega_ba^{\dag}a.
\end{eqnarray}
${\sigma}_z=|L\rangle\langle{L}|-|{R}{\rangle}{\langle}R|$ and
$\sigma_x=|L{\rangle}{\langle}R|+|R{\rangle}{\langle}L|$, where
$|L\rangle$ ($|R\rangle$) is the state of electron occupying left
(right) dot. $\varepsilon$ means the energy level mismatch between
quantum dots which can be tuned by gate voltage. $\Delta$ is the
corresponding tunneling element between two quantum dots. $g$ is
the dots-resonator (e-ph) coupling strength. $\omega_b$ denotes the
quantized mode of the resonator and $a^\dag$ $(a)$ depicts the
corresponding phonon creator (annihilator).

The DQD is connected with two electrodes, described by
$H_{\mathrm{Lead}}=\sum_{j=\{L,R\},k}\epsilon^j_k{c^\dag_{jk}}c_{jk}$
with $\epsilon^{j}_k$ the electronic energy with momentum $k$ in $j$
electrode, and $c^{\dag}_{jk}(c_{jk})$ creating (annihilating) the
corresponding electrons. The Hamiltonian of electrode coupling parts
is given as $V_{\mathrm{DL}}=\sum_{j=\{L,R\},k}{\eta}^j_k|j{\rangle}
{\langle}0 |c_{jk} +\textrm{H.c.}$, where ${\eta}^{j}_k$ is the
coupling strength between DQD and $j$ electrode, and $|0\rangle$
denotes the empty electron state of DQD. The resonator is damped by
external thermal environment,
$H_{\mathrm{E}}=\sum_k\omega_kb^\dag_kb_k$ with $\omega_k$ the
phonon frequency and $b^\dag_k(b_k)$ the corresponding creator
(annihilator). The damping Hamiltonian is described by
$V_{\mathrm{RE}}=\sum_k\lambda_k{a^\dag}b_k+\textrm{H.c.}$, where
$\lambda_k$ denotes the coupling of resonator and the thermal
environment with phonon in momentum $k$. Without loss of generality,
we set $\hbar=1, k_B=1$, $e=1$ and $\omega_b=1$ as the energy unit.

\subsection{Generalized QME and FCS}
To measure fluctuations of electron current through the right
electrode, we add the counting term to the whole Hamiltonian
as~\cite{Esposito}:
\begin{eqnarray}~\label{hwhole}
H_{\chi}&=&e^{\frac{i}{2}{\chi}N_R}He^{-\frac{i}{2}{\chi}N_R} \nonumber \\
&=&H_S+H_{\mathrm{Lead}}+H_{\mathrm{E}}+V_{\mathrm{RE}}+V_{\mathrm{DL}}(\chi),
\end{eqnarray}
where $V_{\mathrm{DL}}(\chi)$ is transformed from the original
$V_{\mathrm{DL}}$, reading:
\begin{eqnarray}
V_{\mathrm{DL}}(\chi)=\sum_{k}\left(e^{-\frac{i}{2}\chi}{\eta}^R_k|R{\rangle}{\langle}0|c_{Rk}+{\eta}^L_k
|L{\rangle}{\langle}0|c_{Lk}\right)+\textrm{H.c.}, \nonumber
\end{eqnarray}
and $N_R=\sum_kc^{\dag}_{Rk}c_{Rk}$ is the electron number operator
in right electrode. $\chi$ is the auxiliary counting parameter in
FCS, appearing in the right coupling part of $V_{\mathrm{DL}}(\chi)$
to count the electron number into the right electrode.

Following the standard procedure, we treat
$V_{\chi}=V_{\mathrm{RE}}+V_{\mathrm{DL}}(\chi)$ as the perturbation. Note here, the couplings between the system $H_S$ and its environments are required to be weak, but not the e-ph coupling $g$ inside $H_S$. Actually, $g$ can be arbitrary strong, compared to the energy scale $\omega_b$.
In this way, under Born-Markov approximation, the
modified QME is derived up to the second order of $V_{\chi}$~\cite{Esposito}:~\label{qme1}
\begin{eqnarray}\label{MQME}
\frac{d\rho^S_{\chi}(t)}{dt}&=&-i[H_S,\rho^S_{\chi}(t)]\nonumber\\
&&-\int^\infty_0d\tau \langle[V_{\chi},[V_{\chi}(-\tau),\rho_{\chi}(t)]_{\chi}]_{\chi}{\rangle}_{\textrm{E,Lead}},
\end{eqnarray}
where
$[A_{\chi},B_{\chi}]_{\chi}=A_{\chi}B_{\chi}-B_{\chi}A_{-\chi}$ for
abbreviation, and
$\rho^S_{\chi}=\textrm{Tr}_{\mathrm{E,Lead}}(\rho_{\chi})$ is the
reduced system density matrix composed of DQD and resonator, by
tracing out degrees of freedom in electrode leads and thermal
environment.${\langle}O{\rangle}_{\textrm{E,Lead}}$ shows thermal
average of $O$ over the thermal reservoir E and two electrode leads. The detail derivation
of Eq.~(\ref{MQME}) can be found in Appendix~\ref{Append_qme}.
Note that $\rho^S_{\chi}$ is a twisted density matrix
by the counting parameter $\chi$. When $\chi=0$, $\rho^S_{\chi}$
reduces to the conventional density matrix.
Therefore, for the system considered here, we finally obtain the
evolution equation of $\rho^S_{\chi}$, as:
\begin{eqnarray}
\frac{d}{dt}{\rho^S_{\chi}}&=&-i[H_S,\rho^S_{\chi}]+L_e[\rho^S_{\chi}]+L_a[\rho^S_{\chi}],
\label{QME}
\end{eqnarray} where $L_e[\rho^S_{\chi}]=$
\begin{multline*}
-\sum_{j=\{L,R\};{n,m}}\bigg({\nu_j}(\epsilon_{nm})
\left(\rho^S_{\chi}|n{\rangle}{\langle}n|S^{\dag}_j|m{\rangle}{\langle}m|S_j 
+\textrm{H.c.}\right)\\
+w_j(\epsilon_{nm})\left(S_j|n{\rangle}{\langle}n|S^{\dag}_j|m{\rangle}{\langle}m|\rho^S_{\chi}+\textrm{H.c.}\right) \\
-w_j(\epsilon_{nm})\left(|n{\rangle}{\langle}n|S^{\dag}_j|m{\rangle}{\langle}m|\rho^S_{\chi}S_j+\textrm{H.c.}\right)e^{-i\chi\delta_{j,R}} \\
-{\nu}_j(\epsilon_{nm})\left(S_j\rho^S_{\chi}|n{\rangle}{\langle}n|S^{\dag}_j|m{\rangle}{\langle}m|+\textrm{H.c.}\right)e^{i\chi\delta_{j,R}}\bigg),
\label{Le}
\end{multline*}
is the contribution from $H_\mathrm{Lead}+V_{\mathrm{DL}}$, and the contribution from $H_\mathrm{E}+V_{\mathrm{RE}}$ reads
\begin{eqnarray}
L_a[\rho^S_{\chi}]=\gamma_b\sum_{\pm}\frac{\bar{n}({\pm}\omega_b)}{{\pm}2}\bigg(2a^{\pm}\rho^S_{\chi}a^{\mp}
-\left\{a^{\pm}a^{\mp},\rho^S_{\chi}\right\}\bigg). \label{La}\nonumber
\end{eqnarray}
Here $S_{j}=|0{\rangle}{\langle}j|$ with $j\in\{L, R\}$, denotes the
annihilation operator pumping electron on dot $j$ into the $j$
electrode, and $S^{\dag}_j=|j{\rangle}{\langle}0|$ is the creator.
$a^+=a^{\dag} (a^-=a)$ creates (annihilates) phonon of the
resonator. $\epsilon_{nm}=\epsilon_n-\epsilon_m$, with $\epsilon_n$
describing the $n$th eigenvalue of $H_S$ and $|n\rangle$ the
corresponding eigenstate.
$w_{j}(\epsilon_{nm})={\Gamma_{j}(\epsilon_{nm})}f_{j}(\epsilon_{nm})/{2}$
and
$\nu_{j}(\epsilon_{nm})=\Gamma_{j}(\epsilon_{nm})[1-f_{j}(\epsilon_{nm})]/{2}$
are the tunneling rates of electrons into and out of the DQD, where
$\Gamma_{j}(\epsilon_{nm})=2\pi\sum_k|{\eta}^{j}_k|^2\delta(\epsilon_{nm}-\epsilon^{j}_k)$
is the spectral function of $j$ electrode and $f_{j}$ depicts
Fermi-Dirac distribution correspondingly.
$\bar{n}(\omega_b)=1/[\exp({\omega_{b}}/{T})-1]$, is the
Bose-Einstein distribution, with $T$ the temperature of the thermal
environment.
$\gamma_b(\omega)=2\pi\sum_k|\lambda_k|^2\delta(\omega-\omega_k)$ is
the spectral function induced by the thermal environment.
$\delta_{j,R}=1$ if $j=R$, otherwise $0$. In the following
calculation, we apply the conventional wide-band limit:
$\gamma_b(\omega)\equiv\gamma_b$,
$\Gamma_{j}(\epsilon_{nm})\equiv\Gamma_{j}$ and consider a large
voltage bias to the electrodes
$({\mu_L}{\gg}{\epsilon,\Delta}{\gg}{\mu_R})$, such that
$f_L(\epsilon_{nm})=1$ and $f_R(\epsilon_{nm})=0$ regardless of the
detail information of $\epsilon_{nm}$. Furthermore, we set
$\bar{n}(\omega_{b})=-1-\bar{n}(-\omega_{b})=0$ for simplicity by
keeping zero temperature of thermal environment, although this
constraint can be released. Considering
$\sum_{n}|n{\rangle}{\langle}n|=1$, the QME with counting parameters
is finally obtained by
\begin{eqnarray}
\frac{d}{dt}\rho^{S}_{\chi}=&&-i[H_S,\rho^{S}_{\chi}]\label{QME2}\\
&&-\frac{\Gamma_L}{2}[S_{L}S^{\dag}_{L}\rho^{S}_{\chi}-2S^{\dag}_{L}
\rho^{S}_{\chi}S_{L}+\rho^S_{\chi}S_{L}S^{\dag}_{L}]\nonumber\\
&&-\frac{\Gamma_R}{2}[S^{\dag}_{R}S_{R}\rho^S_{\chi}-2S_{R}
\rho^{S}_{\chi}S^{\dag}_{R}e^{i{\chi}}+\rho^{S}_{\chi}S^{\dag}_{R}S_{R}]\nonumber\\
&&+\frac{{\gamma}_{b}}{2}[-a^{\dag}a\rho^{S}_{\chi}+2a\rho^{S}_{\chi}a^{\dag}-
\rho^{S}_{\chi}a^{\dag}a].\nonumber
\end{eqnarray}
When $\chi=0$, Eq.~(\ref{QME2}) reduces to the same equation as Eq.~(2) in
Ref.~\cite{Lambert1} and Eq.~(9) in Ref.~\cite{Brandes2}.


Following Ref.~\cite{Esposito}, the moment generating function is
obtained by $\mathcal{G}(\chi,t)=\textrm{Tr}[{\rho}^S_{\chi}(t)]$. The $k$-th
order of charge fluctuations in the right electrode is derived as
${\langle}({\Delta}N_R)^k{\rangle}=\left.\frac{{\partial}^k\mathcal{G}(\chi,t)}{{\partial}(i{\chi})^k}\right|_{{\chi}=0}$,
where ${\Delta}N_R=N_R(t)-N_R(0)$. Then by defining the cumulant generating
function of currents:
\begin{eqnarray}~\label{cgf}
\mathcal{Z}(\chi)=\lim_{t\rightarrow\infty}\frac{1}{t}\ln\left[\mathcal{G}({\chi},t)\right],
\end{eqnarray}
all cumulants of current fluctuations for FCS can be deduced
straightforwardly as
\begin{eqnarray}
\mathcal{I}^{(k)}=\left.\frac{\partial^k\mathcal{Z}(\chi)}{{\partial}(i{\chi})^n}\right|_{\chi=0}.\label{I}
\end{eqnarray}
Fano factor is specified as the ratio of shot noise $S(0)/2=\mathcal{I}^{(2)}$ and current $I=\mathcal{I}^{(1)}$:
\begin{eqnarray}~\label{ff}
\frac{S(0)}{2eI}=\frac{\mathcal{I}^{(2)}}{\mathcal{I}^{(1)}}.\label{S}
\end{eqnarray}

The crucial step to detect current fluctuations, as we shall see, is
to derive the cumulant generating function shown in Eq.~(\ref{cgf}).
In weak e-ph interactions \cite{Lambert1,Brandes2}, current cumulants have been studied under the Fock space of phonons,
by considering the perturbation approximation of the e-ph coupling strength.
For strong e-ph interaction, a huge truncation number of phonons should
be considered to converge the results in numerics, resulting in tough calculations.
Hence it is very challenging to deal with this system by using the conventional Fock
states. While the approach of extended coherent states surmounts such
drawbacks. This method has been successfully applied to Dicke model, spin
boson model and quantum entanglement dynamics~\cite{Chen1,Chen2}.

From the definition, the coherent phonon state is composed
by superpositions of infinite Fock states. Therefore, the finite coherent states already include
infinite phonons in Fock space, which makes the coherent phonon
basis overcomplete. This overcomplete property renders us a rapid
convergence.
We apply this method in the following
by optimally choosing an effective displacement of the resonator
compared to the position in the absence of e-ph interaction. Then, a new
coherent state basis of phonon can be constructed to decouple the
e-ph interaction. As a result, it is expected that coherent states approach is
more effective than the Fock states one, and can deal with arbitrary strong e-ph coupling strengths.

\subsection{Coherent phonon states approach}

We firstly expand the system density operator under electrons
occupation states, and arrange the elements of interest as one
column:
$\bar{\rho}=(\bar\rho_{00},\bar\rho_{LL},\bar\rho_{RR},\bar\rho_{R
L},\bar\rho_{L R})^T$, where
$\bar\rho_{ij}={\langle}i|\rho^S_\chi|j\rangle$, with $|i(j)\rangle$
belonging to $\{|0\rangle,|L\rangle,|R\rangle\}$. Then the evolution
is expressed in a clear way
\begin{eqnarray}
\frac{d}{dt}{\bar{\rho}}&=&\textbf{M}{\bar{\rho}},
\end{eqnarray}
where $\textbf{M}=\textbf{P}+\textbf{D}[\cdot]+L_a[\cdot]$. Here,
$L_a[\cdot]$, is the same as the last right term of Eq.~(\ref{QME2}),
with $\rho^S_\chi$ replaced by $\bar{\rho}$. For $\textbf{P}$, it
is obtained as
\begin{equation}
\textbf{P}=
\begin{pmatrix}
-\Gamma_L & 0 & \Gamma_Re^{i\chi} & 0 & 0\\
\Gamma_L & 0 & 0 & -i\Delta & i\Delta\\
0 & 0 & -\Gamma_R & i\Delta & -i\Delta\\
0 & -i\Delta & i\Delta & -\frac{\Gamma_R}{2}+i\epsilon & 0\\
0 & i\Delta & -i\Delta & 0 & -\frac{\Gamma_R}{2}-i\epsilon
\end{pmatrix},
\end{equation}
which is traditionally used to describe electron transport in the
absence of e-ph interaction~\cite{GK1}. It is contributed by the
first three terms in the right hand side of Eq.~(\ref{QME2}), where
in the first term $-i[H_S, \rho^S_\chi]$, only the pure electron
part $\varepsilon\sigma_z/2+\Delta\sigma_x$ in $H_S$ is included.

The contributions of phonon related parts
$\omega_ba^{\dag}a+g\sigma_z(a^{\dag}+a)$ in $H_S$ in the first
right term of Eq.~(\ref{QME2}) leads to $\textbf{D}[\cdot]$. To get
the expression of $\textbf{D}[\cdot]$, people usually treat the weak
e-ph coupling as the perturbation term. However, as we will show, if
we jump out of the normally used Fock states of phonon, and use a
modified phonon basis, the e-ph coupling can be decoupled
non-perturbatively.

For the case of one electron occupying the $L$ quantum dot,
$\omega_ba^{\dag}a+g\sigma_z(a^{\dag}+a)$ is specified as
$\omega_ba^{\dag}a+g(a^{\dag}+a)$. Then we can define the modified
phonon creator and annihilator:
\begin{eqnarray}
A^{\dag}=a^{\dag}+\alpha, \quad A=a+\alpha, \quad\text{with}\quad
\alpha=\frac{g}{\omega_b}, \label{cpo}
\end{eqnarray}
which are naturally born to decouple the e-ph interaction term by
changing $\omega_ba^{\dag}a+g(a^{\dag}+a)$ to the expression
$\omega_bA^{\dag}A-{g^2}/{\omega_b}$. $\alpha={g}/{\omega_b}$ is
interpreted as the effective displacement deviation of the resonator
induced by e-ph interaction under specified electron state, compared
to the position in the absence of e-ph coupling. Similarly, when one
electron occupies the $R$ quantum dot,
$\omega_ba^{\dag}a+g\sigma_z(a^{\dag}+a)$ is finally decoupled to
$\omega_bB^{\dag}B-{g^2}/{\omega_b}$ by importing
$B^{\dag}=a^{\dag}-\alpha$ and $B=a-\alpha$.
It is clear that with
the help of the modified operators having optimally chosen
displacement $\alpha={g}/{\omega_b}$, the decomposition of e-ph
interaction has been achieved for arbitrary values of $g$, which is
impossible in the conventional Fock space.

Recall $\mathbf{D}[\cdot]$ is derived from a part of $-i[H_S, \rho^S_\chi]$, that is, from $-i[\omega_ba^{\dag}a+g\sigma_z(a^{\dag}+a), \rho^S_\chi]$, when $\sigma_z$ subjected to the electron vacuum state $|0{\rangle}$, one can easily
find $\frac{d}{dt}{\bar{\rho}}_{00}=-i{\omega_b}[a^{\dag}a,{\bar{\rho}}_{00}]$,
similarly as well as for state $|L{\rangle}$,
$\frac{d}{dt}{\bar{\rho}}_{LL}=-i{\omega_b}[A^{\dag}A,{\bar{\rho}}_{LL}]$, and so on.
Thus along this direction, we obtain $\textbf{D}[\bar{\rho}]$ in the new representation:
\begin{eqnarray}
\textbf{D}[\bar{\rho}]=-i{\omega_b}{\mathbf{Diag}}
\begin{pmatrix}
[a^{\dag}a,\bar\rho_{00}]\\
[A^{\dag}A,\bar\rho_{LL}]\\
[B^{\dag}B,\bar\rho_{RR}]\\
B^{\dag}B\bar\rho_{RL}-{\bar\rho_{RL}}A^{\dag}A\\
A^{\dag}A\bar\rho_{LR}-{\bar\rho_{LR}}B^{\dag}B
\end{pmatrix},
\end{eqnarray}
where ${\mathbf{Diag}}\big(\vdots\big)$ means all elements are in
the diagonal positions.

Accordingly, we have electron-states-specified coherent phonon
basis:
\begin{eqnarray}
|n\rangle_0&=&\frac{(a^{\dag})^n}{\sqrt{n!}}|0\rangle_0, \label{cps1}\\
|n\rangle_L&=&\frac{(A^{\dag})^n}{\sqrt{n!}}|0\rangle_L, \label{cps2}\\
|n\rangle_R&=&\frac{(B^{\dag})^n}{\sqrt{n!}}|0\rangle_R, \label{cps3}\\
\text{with}&\quad&
|0\rangle_{L(R)}=e^{-{\alpha^2}/{2}}e^{\mp{\alpha}a^\dag}|0\rangle_0,
\nonumber
\end{eqnarray}
where $|n\rangle_i$ denotes the $n$th exciting phonon state, specified by
the electron state $|i\rangle$ with $i=\{0,L,R\}$. $|0{\rangle}_{0,L,R}$ are the
ground states of the resonator depicted by ${\omega_b}a^{\dag}a$,
${\omega_b}A^{\dag}A$, ${\omega_b}B^{\dag}B$ for different electrons
occupation states. It is straightforward to verify that after the
effective displacement shift $\pm\alpha$, $A^\dag (A)$ with
$|n\rangle_L$ and $B^\dag (B)$ with $|n\rangle_R$ follow the same
creation (annihilation) physics and the same calculation rules as the
conventional un-shifted operators $a^\dag (a)$ with $|n\rangle_0$.

Then, we can insert corresponding coherent phonon states to wholly
expand $\bar{\rho}$ in the form
\begin{eqnarray}
\rho=({\rho_{00}},{\rho_{LL}},{\rho_{RR}},{\rho_{RL}},{\rho_{LR}})^T,\nonumber
\end{eqnarray}
with
\begin{eqnarray}
{\rho_{ji}}=\left(
  \cdots, _j{\langle}n|\bar{\rho}_{ji}|m{\rangle}_i, \cdots
\right),
\end{eqnarray}
where the index $i (j)$ denoting the possible electron state $\{0,
L, R\}$ and $|n\rangle_j, |m\rangle_i$ being the corresponding
phonon states. Therefore, we can finally derive Eq.~(\ref{QME2})
completely as a master equation:
\begin{eqnarray}
\frac{d}{dt}{\rho}&=&\textbf{G}{\rho}, \label{ME}
\end{eqnarray}
where \textbf{G} is the transfer matrix containing counting
parameters. The details of the matrix elements of the above master
equation are described in Appendix~\ref{Append}. The
solution of the above evolution equation reads $\rho(t)=
e^{\mathbf{G}t}\rho(0)$. In the long time limit, the evolution is
dominated by the eigenvalue of $\textbf{G}$ with the largest real
part, ${\lambda_0(\chi)}$, such that $\rho(t)\approx
e^{{\lambda_0(\chi)}t}\rho(0)$~\cite{Ren}. Note when $\chi=0$, $\lambda_0(\chi)$ reduces to zero,
which corresponds to the steady state of the dynamics without counting parameters. Therefore, the eigenvalue
of $\textbf{G}$ with the largest real part is the cumulant
generating function at the steady state,
$\mathcal{Z}(\chi)={\lambda_0(\chi)}$. Then, we can apply
Eq.~(\ref{I}) and (\ref{S}) to investigate the quantum transport
properties, by numerical differentiation.

In previous studies~\cite{Lambert1,Brandes2}, current cumulants have
been derived under bosonic Fock space. However, in that space the
e-ph interaction could only be tuned in weak e-ph coupling strength.
The main reason is as the e-ph coupling
strength $g$ is weak, the e-ph interaction term
$g\sigma_z(a^{\dag}+a)$, represented in $\textbf{D}[\cdot]$, can be
treated as the perturbation. As a result, it is straightforward to
solve Eq.~(\ref{QME2}) for the small value of e-ph coupling $g$ under
Fock space of phonon with convergent results by truncating phonon
Hilbert space, shown in Ref. ~\cite{Lambert1,Brandes2}.
However, when $g$ is not small, like the usual real situations,
the perturbation approximation of the e-ph interaction is no longer
valid. Tremendous high phonon states will be excited by the strong
e-ph coupling, in which case calculations become tough
from the perspective of Fock states.

The advantage of coherent phonon states is that the modified creator
and annihilator favors the displacement of resonator induced by e-ph
coupling which in turn decouples the e-ph coupling
non-perturbatively, as we show in Eq.~(\ref{cpo}) and its following
discussions. Moreover, from Eq.~(\ref{cps2}) and (\ref{cps3}), it is
clear that the coherent phonon state actually is the superposition
of infinite number of Fock states, which makes the coherent phonon
basis overcomplete. This overcomplete property renders us rapid
convergent calculations under the optimally chosen displacement
$\alpha=g/\omega_b$. Therefore, it is natural to choose coherent
phonon states as the proper basis to investigate the effects of e-ph
interaction on quantum transports.

Further, we would like to point out that the extended coherent phonon states approach shares the same physics with the polaron (canonical, Lang-Firsov) transformation \cite{Mahanbook} that they both consider the displaced phonon basis. While for the detail mathematical treatment, they are different. The polaron transformation method is applied to decouple the e-ph interaction, but makes the system-electrode tunneling terms more complex, by adding a cloud of phonons to the operators of electrons, so-called ``dressed'' states \cite{Mahanbook}. The extended phonon states approach also decouples the e-p coupling; furthermore, it conserves the simplicity of the system-electrode tunneling term. Hence, it makes the practical calculations of the results (current, cumulants etc.) efficient and comprehensible. As a result, we use the extended coherent phonon states approach to investigate the quantum transport properties in the present work.

\begin{figure}
\centering
\resizebox{0.38\textwidth}{!}{\includegraphics{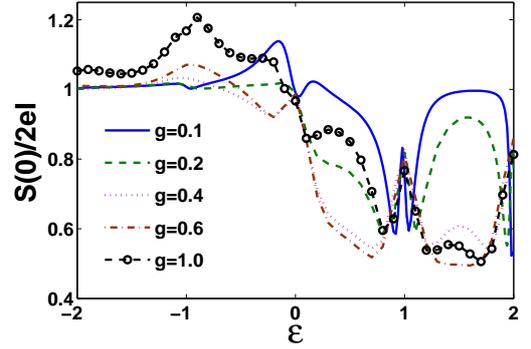}}
\caption{(Color online) Fano factor ${S(0)}/({2eI})$ versus DQD energy
gap $\epsilon$ with various $g$. The other parameters are given as $\Delta=0.1$,
$\Gamma_L=0.1$, $\Gamma_R=0.001$ and $\gamma_b=0.01$. $\omega_b=1$
is set as the energy unit.}\label{Figure}
\end{figure}

\section{Results and discussions}


In Fig.~\ref{Figure}, we first show Fano factor (see Eq.~(\ref{ff})) as comparisons with previous results
given in Ref.~\cite{Lambert1} to validate our method.
When the DQD-resonator coupling is weak $(g=0.1,0.2,0.4)$, we find
the results obtained by our method are the same as those in
Ref.~\cite{Lambert1}. Moreover, our method with coherent phonon
states can also work at strong e-ph coupling regime $(g=1.0)$, where
the previous method with Fock states fails.

In the following, with the help of coherent phonon states, we are capable to scrutinize the DQD's transport properties mediated by
e-ph coupling in a full range of strengths. We focus on analyzing the first three cumulants of probability distribution
of electron current, though higher order cumulants are also available.

\begin{figure}
\resizebox{0.42\textwidth}{!}{\includegraphics{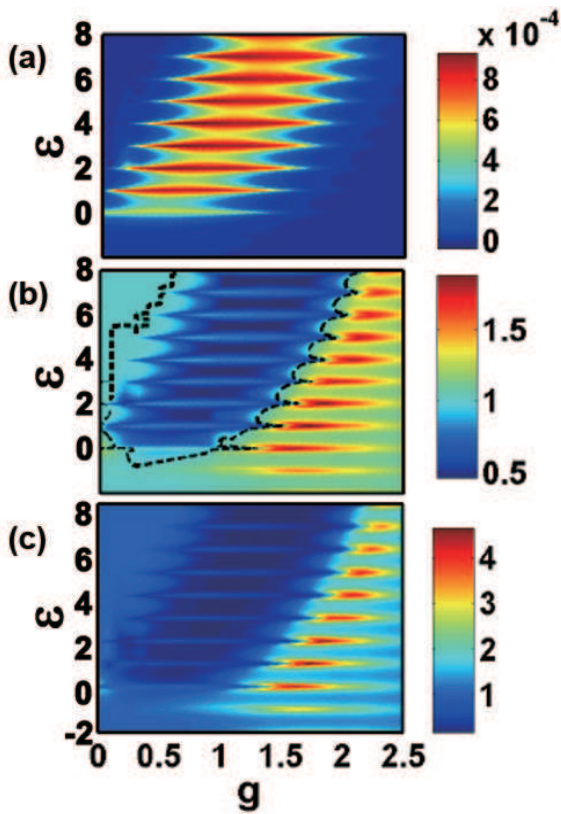}}\vspace{0.5cm}
\caption{(Color online) Current ($\mathcal{I}^{(1)}$), Fano factor
(${\mathcal{I}^{(2)}}/{\mathcal{I}^{(1)}}$) and RTOC
(${\mathcal{I}^{(3)}}/{\mathcal{I}^{(1)}}$) are numerically
calculated in (a), (b) and (c) under parameters of $g$ and
$\varepsilon$. Dashed black line in (b) is used to signify Poissonion transport with Fano
factor equal to $1$. The other parameters are given as $\Delta=0.1$,
$\Gamma_L=0.1$, $\Gamma_R=0.001$, $\gamma_b=0.01$ and $\omega_b=1$.
}\label{Figure2}
\end{figure}

In Fig.~\ref{Figure2}(a), by tuning the energy level mismatch $\varepsilon$ and e-ph coupling $g$, we find there exist multiple resonance islands of current, where $\mathcal{I}^{(1)}$ suddenly becomes
large under the integer relation $\varepsilon=k\omega_b$,
$k\in\mathbb{N}$, as is also discovered in Refs.~\cite{Brandes2,Armour}. This integer relation indicates that the electron
tunneling is strongly assisted by $k$-phonon excitations from the
mechanical resonator when the energy gap of two quantum dots is equal to the energy of $k$ phonons. That is, the tunneling from quantum dot $L$ to $R$ will be significantly enhanced if the excess energy can excite integer number of phonons of the resonator's vibration, which are then absorbed by the zero temperature thermal
environment.

The positive integer resonance is, however, not always satisfied in regimes of large energy mismatch of DQD $(\varepsilon\gg{\omega_b})$ and weak
e-ph coupling $(g\ll{\omega_b})$ where it is insufficient to excite
extra phonons, or in regimes of
extremely strong coupling $(g\gg{\omega_b})$ where e-ph coupling
plays the role of scattering and strongly represses the electron
current. For negative $\epsilon$, the current is drastically suppressed
by the fact that the resonator has the ability only to emit phonons
to the thermal environment of zero temperature. If we increase the
temperature of the environment, the resonance peak emerging at
negative integer values of $\epsilon$ will be observed. We note that in a different setup of triple quantum dots \cite{Dominguez}, similar integer relation for current is exhibited, but as dips (repressions) rather than peaks (enhancements).

Fano factor shown in Fig.~\ref{Figure2}(b), is more complicated than the
current. We separate the plot into two sub-regimes by the dashed black
line, which notifies Poissonion transport. In the regime surrounded
by this line, Fano factor drops into sub-Poisson regime, where the
coupling strength $g$ is moderate~\cite{Lambert1}. The shrink of
Fano factor mainly results from the dramatic increasing of current
induced by e-ph excitations. When $g$ becomes strong, Fano factor
rises to super-Poisson regime due to the repression of currents. 
Furthermore, resonance enhancements of Fano factor at integer
$\varepsilon$ are clearly observed in both regimes of positive
and negative $\varepsilon$ with strong $g$.

Fig.~\ref{Figure2}(c) shows the renormalized third order cumulant (RTOC),
which should not be ignored to signify high order transport
fluctuations~\cite{Flindt,Flindt2}. For moderate $g$, RTOC is rather
small due to resonances of the current. While $g$ reaches strong regime, RTOC is strengthened,
mainly due to the repression of currents shown in Fig.~\ref{Figure2}(a).
The multi-phonon-resonance-induced RTOC enhancement still
appears, which is similar to what happens for shot noise. 

\begin{figure}
\resizebox{0.45\textwidth}{!}{\includegraphics{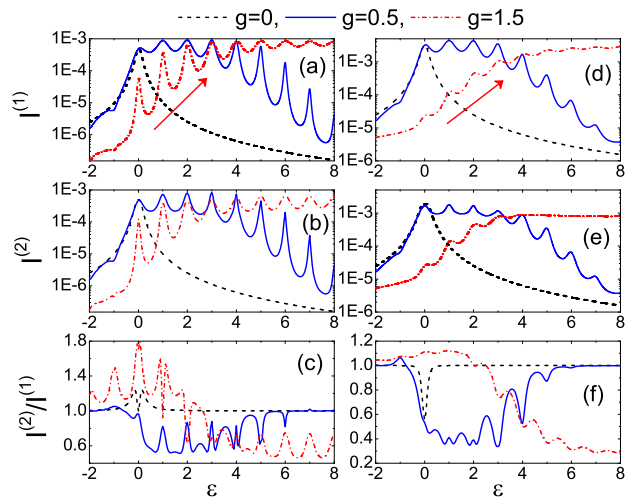}}\vspace{0.5cm}
\caption{(Color online) Current $\mathcal{I}^{(1)}$, shot noise $\mathcal{I}^{(2)}$ and Fano factor $\mathcal{I}^{(2)}/\mathcal{I}^{(1)}$ as a function of
$\varepsilon$ for different $g$. Asymmetric tunneling rates
condition: $\Delta=0.1$, $\Gamma_L=0.1$, $\Gamma_R=0.001$ and
$\gamma_b=0.01$ for (a), (b) and (c); Symmetric tunneling rates
condition: $\Delta=0.1$, $\Gamma_L=0.01$, $\Gamma_R=0.01$ and
$\gamma_b=0.05$ for (d), (e) and (f). 
}\label{Figure3}
\end{figure}

To understand the complicated behaviors of Fano factor, we detail
the dependence of current fluctuations on DQD energy gap
$\varepsilon$ in Fig.~\ref{Figure3}.
Peaks appear at $\varepsilon=k\omega_b$ for both current and shot
noise, resulting from the integer phonon resonance. When $g$ is
moderate (exemplified by $g=0.5$), oscillatory decays of both
current and shot noise in Fig.~\ref{Figure3}(a)(b) and (d)(e) are
visible for $\varepsilon>4.0$ while Fano factor rises gradually from
sub-Poisson regime to $1.0$ with oscillations, see
Fig.~\ref{Figure3}(c)(f). This can be interpreted that for large
$\varepsilon$ compared to $g$, the e-ph coupling is not able to
assist the electron transport and tunneling events between quantum
dots are seldom, so that successive transmissions of electrons are
almost uncorrelated. As a result, the transport dynamics approaches
Poissonion regime such that $S(0) \approx 2eI$. When $g$ is strong
(exemplified by $g=1.5$), current and shot noise are lifted as
$\varepsilon$ increases (see Fig.~\ref{Figure3}(a)(b)), indicated by
the arrows. It suggests that multi-phonon excitations are definitely
favored in large e-ph coupling $g$ regime to enhance both the current
and shot noise. Though not depicted here, we still observe that
current and corresponding shot noise will reach the maximum and then
decreases when we increase $\varepsilon$ to extremely large values.

Although the phonon-assisted tunneling that is excited by e-ph
coupling is not expected at the negative integer value of
$\varepsilon$ at zero temperature, the resonance peaks are still
observed in Fig.~\ref{Figure3}(c)(f) for Fano factor. These peaks
result from the dips of current at the negative integer
$\varepsilon$, shown in Fig. \ref{Figure3}(a)(d). The dips are due
to the interference between the elastic and inelastic transmission
scattered by the e-ph coupling, instead of opening new tunneling
channels by e-ph coupling~\cite{Galperin}.

\begin{figure}
\resizebox{0.45\textwidth}{!}{\includegraphics{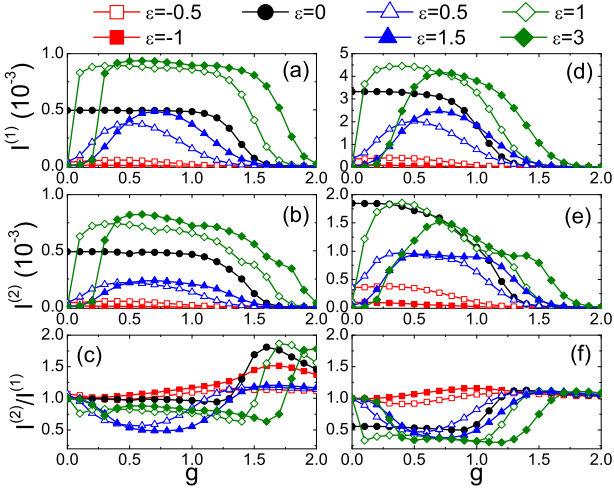}}\vspace{0.5cm}
\caption{(Color online) Current $\mathcal{I}^{(1)}$, shot noise $\mathcal{I}^{(2)}$ and Fano factor $\mathcal{I}^{(2)}/\mathcal{I}^{(1)}$ as a function of
$g$ for some representative $\varepsilon$. Asymmetric tunneling
rates condition: $\Delta=0.1$, $\Gamma_L=0.1$, $\Gamma_R=0.001$ and
$\gamma_b=0.01$ for (a), (b) and (c); Symmetric tunneling rates
condition: $\Delta=0.1$, $\Gamma_L=0.01$, $\Gamma_R=0.01$ and
$\gamma_b=0.05$ for (d), (e) and (f).
}\label{Figure4}
\end{figure}

The dependencies of current, shot noise and Fano factor on $g$ are
detailed in Fig.~\ref{Figure4}, with four typical transport
behaviors exemplified therein:

(1) The first kind behavior appears at $\varepsilon<0$ (exemplified by $\varepsilon=-0.5, -1$), where current and shot noise are both
rather small in the whole regime of $g$. This is understandable
since in this regime electrons in DQD are severely localized, and
there is no feedback of phonons from noise environment with zero
temperature. As a result, electrons cannot effectively transfer
through DQD by phonon-assisted tunneling.

(2) $\varepsilon=0$ depicts the second kind behavior. When $g\lesssim1.0$, current, shot noise and Fano factor are only slightly affected, which show that for the DQD with two degenerate energy levels, the transport properties are robust for weak e-ph couplings. When $g$ increases further, the scattering effect of e-ph coupling becomes significant and the current decreases. In the absence of e-ph coupling and $\varepsilon=0$, the current can be
described~\cite{Nazarov1} by
\begin{eqnarray}
\mathcal{I}^{(1)}=\frac{{\Delta}^2\Gamma_R}{\Delta^2(2+\Gamma_R/\Gamma_L)+\Gamma_R^2/4+\varepsilon^2}.
\end{eqnarray}
Consequently, for the case of asymmetric tunneling rates, $\mathcal{I}^{(1)}\approx{\Gamma_R}/{2}=0.0005$, while for the symmetric tunneling case
$\mathcal{I}^{(1)}\approx{\Gamma_R}/{3}=0.0033$, consistent with what are shown in
Fig. \ref{Figure4}(a) and (d). In Refs.~\cite{Elattari,Brandes3}, the exact result of
Fano factor is obtained without considering e-ph coupling ($g=0$):
\begin{equation}
1-8{\Delta^2}\frac{4\varepsilon^2(\Gamma_R/\Gamma_L-1)+
3\Gamma_R^2+\Gamma_R^3/\Gamma_L+8\Gamma_R/\Gamma_L\Delta^2}
{[\Gamma_R^2+4\varepsilon^2+4\Delta^2(\Gamma_L/\Gamma_R+2)]^2}.
\end{equation}
For $\varepsilon=0$, we then can approximately simplify Fano factor as $1$ for asymmetric tunneling case $\Gamma_R\ll\Gamma_L$ and as $\frac{5}{9}$ for symmetric tunneling case $\Delta\gg\Gamma_R=\Gamma_L$. These coincide with the simulation results shown in  Fig. \ref{Figure4}(c) and (f), where Fano factor is nearly $1.0$ for the asymmetric case, indicating the Poissonion nature while it is strongly suppressed around $0.5$ when $g\lesssim1.0$ for the symmetric tunneling.

(3) The third kind behavior is for positive but non-integer $\varepsilon$ (exemplified by $\varepsilon=0.5, 1.5$). As shown in Fig.~\ref{Figure4}(a) and (d), these two curves of currents first increase when $g$ increases as a consequence of phonon-assisted tunneling. An then, when the e-ph coupling strength $g$ is further increased, the currents decrease because of the phonon-scattering-induced localization.
In Fig.~\ref{Figure4}(b)(e), shot noise changes almost synchronously with currents, however, with smaller amplitude variations
compared to currents. Hence, Fano factor experiences from
sub-Poisson to Poisson regime, gaining a global minimum around
$0.5$. 

(4) Positive integer $\varepsilon$ (exemplified by $\varepsilon=1, 3$) depicts the forth kind behavior, that is, the resonant electron tunneling regime. The e-ph coupling
first plays a constructive role to assist electron tunneling.
While, as the coupling strength $g$ increases, the resonator's vibration scatters
electrons and suppresses the current dramatically. This is similar to the non-integer $\varepsilon$ cases and the difference is that the maximum of the resonant cases is more like a platform. Generally, near the resonances, one can write \cite{Haugbook}:
\begin{eqnarray}
\mathcal{I}^{(1)}=\int\frac{d\varepsilon}{2\pi} \frac{\Gamma_L\Gamma_R}{(\varepsilon-k\omega_b)^2+(\Gamma_L+\Gamma_R)^2/4}[f_L-f_R].
\end{eqnarray}
This can be successively approximated as $\mathcal{I}^{(1)}\approx \frac{\Gamma_L\Gamma_R}{\Gamma_L+\Gamma_R}$ in the weak $\Gamma_{L(R)}$ limit, since the Lorentz function reduces to the delta function.
Therefore, for asymmetric tunneling case $\Gamma_R\ll\Gamma_L$, the resonant current nearly reaches the maximum $\mathcal{I}^{(1)}=\Gamma_R$ and for symmetric tunneling case $\Gamma_R=\Gamma_L$, the resonant current nearly reaches the maximum $\mathcal{I}^{(1)}={\Gamma_R}/{2}$, which are consistent with our calculations shown in Fig.~\ref{Figure4}(a) and (d).
Furthermore, different from the non-integer $\varepsilon$ cases, for integer $\varepsilon$ Fano factor exhibits two local minimums at its minimum basin in the sub-Poisson regime, as shown in Fig.~\ref{Figure4}(c) and (f).

\begin{figure}
\resizebox{0.45\textwidth}{!}{\includegraphics{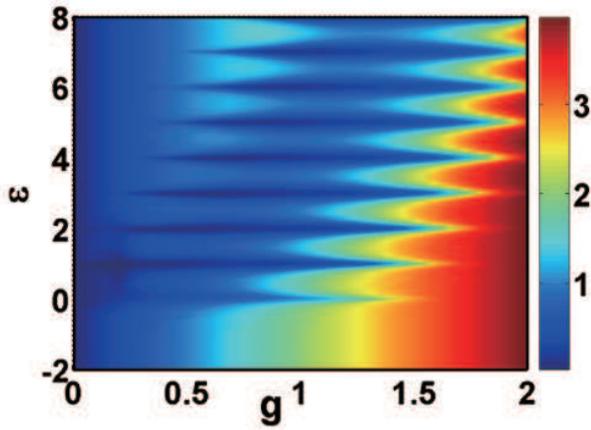}}\vspace{0.5cm}
\caption{(Color online) ${\langle}\sigma_z(a^\dag+a){\rangle}$ as a function of $g$
and $\varepsilon$. Other parameters are $\Delta=0.1$,
$\Gamma_L=0.1$, $\Gamma_R=0.001$, $\gamma_b=0.01$. }\label{Figure5}
\end{figure}

In Fig.~\ref{Figure5}, we show the expectation of the e-ph interaction term,
${\langle}\sigma_z(a^\dag+a){\rangle}$, to exam the role of e-ph interaction in the transport. ${\langle}\sigma_z(a^\dag+a){\rangle}=\langle|L{\rangle}{\langle}L|(a^{\dag}+a){\rangle}-\langle|R{\rangle}{\langle}R|(a^{\dag}+a){\rangle}$ describes the discrepancy of resonator displacement's contributions on electron occupancies on the left and right quantum dots.
Under resonance conditions $(\varepsilon=k\omega_b)$, the discrepancy
${\langle}\sigma_z(a^\dag+a){\rangle}$ is rather small, compared to off-resonance
cases with the same $g$, which indicates the electron occupation probabilities on the two quantum dots are almost equivalent. That is, the phonon emission of the resonator into the thermal environment assists electron transferring from the left quantum dot to the right one, rather than blocks electron transport. This is consistent with multi-phonon-assisted resonance discussed in Fig.~\ref{Figure2} and~\ref{Figure3}.

However, when the e-ph coupling $g$ becomes strong, the discrepancy ${\langle}\sigma_z(a^\dag+a){\rangle}$
becomes significant for given $\varepsilon$, which implies the phonon assistance mechanism is depressed. The e-ph interaction no longer assists electron tunneling, but plays the role of scattering.
Hence, the electron is mainly localized in the left quantum dot (indicated by the significantly positive value of ${\langle}\sigma_z(a^\dag+a){\rangle}$) due to the e-ph scattering, which
results in the repression of the electron current, as shown
in Figs.~\ref{Figure2} and~\ref{Figure4}.

\section{Summary}

To conclude, we have applied the modified Born-Markov quantum master equation to
investigate full counting statistics of electron transport through the double quantum dots coupled with a nanomechanical resonator. Particularly, with the help of coherent phonon states method, we are capable to non-perturbatively treat the e-ph coupling of arbitrary strong strength, with excellent convergence.
We have shown the first three order current fluctuations in detail to
signify the non-Poissonion electron transport features in different dynamics regimes.
Conditional phonon-assisted resonant tunneling has been observed under
the positive integer relation of $\varepsilon=k\omega_b, k\in\mathbb{N}$. We have further found that, in strong e-ph coupling regime, multi-phonon excitations are favored
such that transport is enhanced with increasing energy gap
$\varepsilon$. Moreover, for positive $\varepsilon$, we have found that
as the strength $g$ increases, the e-ph coupling first plays a constructive role to assist the transport, and then plays the role of scattering and strongly represses the transport. Finally, we have studied the expectation value of DQD-resonator interaction
to show the difference of contributions between quantum
dots, which complements out understanding on multi-phonon-assisted electron tunneling and phonon-scattering-induced electron localization in strong e-ph
regimes.

Though not fully exploited yet, the extended coherent phonon states method may
be extended to directly investigate dissipative DQD system with
multiple-phonon-electron coupling~\cite{Chen2}. We expect our method
can be applied to other open quantum systems beyond weak e-ph
interaction constraints, to uncover possible novel many-body quantum effects
induced by strong e-ph interactions, like the avalanche type of transport \cite{Koch}.

\section{Acknowledgements}
We thank Dr. Y. Y. Zhang for helpful discussions. This work (C.W. and Q.H.C.)
is supported in part by National Natural Science
Foundation of China under Grant No. 10974180, National Basic Research Program of China under Grant Nos. 2011CBA00103 and  2009CB929104.

\appendix
\section{Derivation of Quantum Master Equation under Counting Field}\label{Append_qme}
The whole Hamiltonian including system and environments under counting field can be
described by $H_{\chi}=H_0+V_{\chi}$ given in Eq.~(\ref{hwhole}). Here $H_0=H_S+H_{\textrm{Lead}}+H_E$ shows noninteracting term
and $V_{\chi}=V_{\textrm{RE}}+V_{\textrm{DL}}(\chi)$ is treated as perturbation one, assuming the interacting between system and
environment/electrodes is weak (not the e-ph coupling inside the system).

Under interacting picture, the motion equation of density matrix from the whole model can be shown as
\begin{eqnarray}~\label{dm1}
\frac{d{\rho^I_{\chi}(t)}}{d{t}}=-i[V_{\chi}(t),\rho^I_{\chi}(t)],
\end{eqnarray}
where $\rho^I_{\chi}(t)=e^{iH_0t}\rho_{\chi}(t)e^{-iH_0t}$ and $V_{\chi}(t)=e^{iH_0t}V_{\chi}e^{-iH_0t}$.
By integrating Eq.~(\ref{dm1}), density matrix can be expressed as
\begin{eqnarray}
{\rho^I_{\chi}(t)}&=&\rho_{\chi}(0)-i\int^{t}_{0}dt'[V_{\chi}(t'),{\rho^I_{\chi}(t')}],\\
&=&\rho_{\chi}(0)-i\int^{t}_{0}dt'[V_{\chi}(t'),{\rho_{\chi}(0)}]\nonumber\\
&&-\int^{t}_{0}dt'\int^{t'}_0d{\tau}[V_{\chi}(t'),[V_{\chi}(\tau),\rho^I_{\chi}(\tau)]].\nonumber
\end{eqnarray}
Hence the density matrix motion equation in interacting picture can be derived by
\begin{eqnarray}~\label{dm2}
\frac{d{\rho^{I}_{\chi}(t)}}{dt}&=&-i[V_{\chi}(t),\rho_{\chi}(0)]\nonumber\\
&&-\int^{t}_{0}d{\tau}[V_{\chi}(t),[V_{\chi}(\tau),\rho^I_{\chi}(\tau)]],\nonumber\\
\end{eqnarray}
Transforming it back to the Schr\"{o}dinger picture, one can find the reduced system density matrix can be expressed by
$\rho^{S}_{\chi}(t)=\textrm{Tr}_\textrm{{E,Lead}}[e^{-iH_0t}\rho^{I}_{\chi}(t)e^{iH_0t}].$
As a result, the motion equation of the reduced system density matrix is shown by
\begin{eqnarray}
\frac{d\rho^S_{\chi}(t)}{dt}&=&-i[H_S,\rho^S_{\chi}(t)]\nonumber\\
&&-\int^\infty_0d\tau{\langle}[V_{\chi},[V_{\chi}(-\tau),\rho_{\chi}(t)]_{\chi}]_{\chi}{\rangle}_{\textrm{E,Lead}}.\nonumber\\
\end{eqnarray}
Here Born approximation is applied to decompose the whole density matrix
as $\rho_{\chi}(t)=\rho^{S}_{\chi}(t)\otimes{\rho_E}\otimes{\rho_{{Lead}}}$.
Markov approximation should be also considered to replace $\rho^{I}_{\chi}(\tau)$ in the right second term
of Eq.~(\ref{dm2}) by $\rho^{I}_{\chi}(t)$ and extend the upper integral limit from $t$
to $\infty$.
Besides, $[A_{\chi},B_{\chi}]_{\chi}=A_{\chi}B_{\chi}-B_{\chi}A_{-\chi}$ and ${\langle}O{\rangle}_{\textrm{E,Lead}}$
shows statistical average of $O$ over thermal environment and leads.

\section{Derivation of Density Matrix Equation}\label{Append}
The evolution Eq.~(\ref{ME}) has five blocks specified by the five
electron states. Let us denote the reduced system density matrix
elements combined with counting field as
\begin{eqnarray}
\rho^{ji}_{n,m}={_j{\langle}}n|{\langle}j|\rho^S_{\chi}|i{\rangle}|m{\rangle}_i.\nonumber
\end{eqnarray}
Then the first block describes the evolution of elements in
${\rho_{00}}$:
\begin{eqnarray}
&&\frac{d}{dt}\rho^{00}_{n,m}=-i\omega_b(n-m)\rho^{00}_{n,m}-{\Gamma_L}\rho^{00}_{n,m}\nonumber\\
&&+\Gamma_R\sum_{k,l}{_0\langle}n|k{\rangle_R}{_R\langle}l|m{\rangle_0}e^{i\chi}\rho^{RR}_{k,l}\nonumber\\
&&-\frac{\gamma_b}{2}[(n+m)\rho^{00}_{n,m}-2\sqrt{(n+1)(m+1)}\rho^{00}_{n+1,m+1}]\nonumber\\
&&+\bar{n}\gamma_b[-(n+m)\rho^{00}_{n,m}+\sqrt{(n+1)(m+1)}\rho^{00}_{n+1,m+1}\nonumber\\
&&+\sqrt{nm}\rho^{00}_{n-1,m-1}].\nonumber
\end{eqnarray}
The second block is for the evolution of ${\rho_{LL}}$:
\begin{eqnarray}
&&\frac{d}{dt}\rho^{LL}_{n,m}=-i\omega_b(n-m)\rho^{LL}_{n,m}
+{\Gamma_L}\sum_{k,l}{_L\langle}n|k{\rangle_0}{_0\langle}l|m{\rangle_L}\rho^{00}_{k,l}\nonumber\\
&&-i\Delta\sum_{k}{_L\langle}n|k{\rangle_R}\rho^{RL}_{k,m}+i\Delta\sum_{k}\rho^{LR}_{n,k}{_R\langle}k|m{\rangle_L}\nonumber\\
&&-\frac{\gamma_b}{2}[(n+m)\rho^{LL}_{n,m}-\alpha\sqrt{n}\rho^{LL}_{n-1,m}\nonumber\\
&&+\alpha\sqrt{n+1}\rho^{LL}_{n+1,m}+\alpha\sqrt{m+1}\rho^{LL}_{n,m+1}\nonumber\\
&&-\alpha\sqrt{m}\rho^{LL}_{n,m-1}-2\sqrt{(n+1)(m+1)}\rho^{LL}_{n+1,m+1}]\nonumber\\
&&+\bar{n}\gamma_b[-(n+m)\rho^{LL}_{n,m}+\sqrt{(n+1)(m+1)}\rho^{LL}_{n+1,m+1}\nonumber\\
&&+\sqrt{nm}\rho^{LL}_{n-1,m-1}].\nonumber
\end{eqnarray}
The third block is for the evolution of ${\rho_{RR}}$:
\begin{eqnarray}
&&\frac{d}{dt}\rho^{RR}_{n,m}=-i\omega_b(n-m)\rho^{RR}_{n,m}
-\Gamma_R\rho^{RR}_{n,m}\nonumber\\
&&+i\Delta\sum_{k}{_L\langle}k|m{\rangle_R}\rho^{RL}_{n,k}-i\Delta\sum_{k}{_R\langle}n|k{\rangle_L}\rho^{LR}_{k,m}\nonumber\\
&&-\frac{\gamma_b}{2}[(n+m)\rho^{RR}_{n,m}+\alpha\sqrt{n}\rho^{RR}_{n-1,m}\nonumber\\
&&-\alpha\sqrt{n+1}\rho^{RR}_{n+1,m}-\alpha\sqrt{m+1}\rho^{RR}_{n,m+1}\nonumber\\
&&+\alpha\sqrt{m}\rho^{RR}_{n,m-1}-2\sqrt{(n+1)(m+1)}\rho^{RR}_{n+1,m+1}].\nonumber
\end{eqnarray}
The fourth block is for the evolution of ${\rho_{R L}}$:
\begin{eqnarray}
&&\frac{d}{dt}\rho^{RL}_{n,m}=i[\varepsilon\rho^{RL}_{n,m}+\Delta\sum_{k}{_R\langle}k|m{\rangle_L}\rho^{RR}_{n,k}\nonumber\\
&&-\Delta\sum_{k}{_R\langle}n|k{\rangle_L}\rho^{LL}_{k,m}+\omega_b(m-n)\rho^{RL}_{n,m}]\nonumber\\
&&-\frac{\Gamma_R}{2}\rho^{RL}_{n,m}-\frac{\gamma_b}{2}[(n+m+4\alpha^2)\rho^{RL}_{n,m}\nonumber\\
&&+3\alpha\sqrt{n+1}\rho^{RL}_{n+1,m}-\alpha\sqrt{n}\rho^{RL}_{n-1,m}\nonumber\\
&&-3\alpha\sqrt{m+1}\rho^{RL}_{n,m+1}-\alpha\sqrt{m}\rho^{RL}_{n,m-1}\nonumber\\
&&-2\sqrt{(n+1)(m+1)}\rho^{RL}_{n+1,m+1}]\nonumber\\
&&+\bar{n}\gamma_b[-(n+m+4\alpha^2)\rho^{RL}_{n,m}\nonumber\\
&&-2\alpha(\sqrt{n+1}\rho^{RL}_{n+1,m}+\sqrt{n}\rho^{RL}_{n,m})\nonumber\\
&&+2\alpha(\sqrt{m+1}\rho^{RL}_{n,m+1}+\sqrt{m}\rho^{RL}_{n,m-1})\nonumber\\
&&+\sqrt{(n+1)(m+1)}\rho^{RL}_{n+1,m+1}+\sqrt{nm}\rho^{RL}_{n-1,m-1}].\nonumber
\end{eqnarray}
So is the last block for ${\rho_{L R}}$:
\begin{eqnarray}
&&\frac{d}{dt}\rho^{LR}_{n,m}=-i[\varepsilon\rho^{LR}_{n,m}+\Delta\sum_{k}{_L\langle}n|k{\rangle_R}\rho^{RR}_{k,m}\nonumber\\
&&-\Delta\sum_{k}{_L\langle}k|m{\rangle_R}\rho^{LL}_{n,k}+\omega_b(n-m)\rho^{LR}_{n,m}]\nonumber\\
&&-\frac{\Gamma_R}{2}\rho^{LR}_{n,m}-\frac{\gamma_b}{2}[(n+m+4\alpha^2)\rho^{LR}_{n,m}\nonumber\\
&&-3\alpha\sqrt{n+1}\rho^{LR}_{n+1,m}-\alpha\sqrt{n}\rho^{LR}_{n-1,m}\nonumber\\
&&+3\alpha\sqrt{m+1}\rho^{LR}_{n,m+1}+\alpha\sqrt{m}\rho^{LR}_{n,m-1}\nonumber\\
&&-2\sqrt{(n+1)(m+1)}\rho^{LR}_{n+1,m+1}]\nonumber\\
&&+\bar{n}\gamma_b[-(n+m+4\alpha^2)\rho^{LR}_{n,m}\nonumber\\
&&+2\alpha(\sqrt{n+1}\rho^{LR}_{n+1,m}+\sqrt{n}\rho^{LR}_{n-1,m})\nonumber\\
&&-2\alpha(\sqrt{m+1}\rho^{LR}_{n,m+1}+\sqrt{m}\rho^{LR}_{n,m-1})\nonumber\\
&&+\sqrt{(n+1)(m+1)}\rho^{LR}_{n+1,m+1}+\sqrt{nm}\rho^{LR}_{n-1,m-1}].\nonumber
\end{eqnarray}
For the inner-product of the coherent phonon states in these
evolution equations, the expression can be deduced by following
Eq.~(\ref{cps1}) to (\ref{cps3}):
\begin{eqnarray}
_{L}{\langle}n|m\rangle_{0}&=&_{0}{\langle}n|m\rangle_{R}=(-1)^{m}D_{nm}(\alpha),
\nonumber \\
_{L}{\langle}n|m\rangle_{R}&=&(-1)^{m}D_{nm}(2\alpha), \nonumber \\
_{R}{\langle}n|m\rangle_{L}&=&(-1)^{m}D_{nm}(-2\alpha), \nonumber \\
\text{with} \quad D_{nm}(x)&=&\sqrt{n!m!} \exp{(-\frac{x^2}{2})}
x^{n+m}\nonumber\\
&&\times \sum_{k=0}^{\min\{n,m\}}\frac{x^{-2k}
(-1)^k}{(n-k)!(m-k)!k!}.\nonumber
\end{eqnarray}

In the practical calculations, the truncation number of phonon
occupation states is set as $\textrm{Ntr}$. Given a specified
$\textrm{Ntr}$, there exist $5(\textrm{Ntr}+1)^2$ equations in all,
where the $5$ characterizes the space dimension of electron states.
Thanks to the overcomplete property of coherent phonon states, the
truncation number of modified phonon space has much shrinked even
when $g$ becomes large. Through our whole work, we set
$\textrm{Ntr}\leq30$. We have checked that this number is enough and
all the results converge with relative errors less than $10^{-5}$.

\end{document}